# Surface Enhanced Circular Dichroism of Oriented Chiral Molecules by Plasmonic Nanostructures

*Weixuan Zhang, Tong Wu, Rongyao Wang and Xiangdong Zhang\**

Beijing Key Laboratory of Nanophotonics & Ultrafine Optoelectronic Systems, School of Physics, Beijing Institute of Technology, Beijing, 100081, China

**ABSTRACT:** We present a rigorous finite element method to calculate circular dichroism (CD) in various systems consisting of nanostructures and oriented chiral molecules with electric quadrupole transitions. The interaction between oriented molecule materials, which are regarded as anisotropic chiral media, and metallic nanostructures has been investigated. Our results show that the plasmon-induced CD is sensitive to the orientations of the molecules. In many cases, the contribution of molecular electric quadrupole transitions to the total CD signal can play a key role. More interesting, we have demonstrated that both the quadrupole- and dipole-based CD signals can be improved greatly by matching the phases for the electromagnetic fields and their gradients at different regions around the nanostructures, which are occupied by the oriented chiral molecules. Different regions might produce CD of opposite sign. When integrating over regions with only one side of the proposed nanostructure, we find that the CD-peak may be nearly hundreds-fold over the case of integrating both sides. We believe that these findings would be helpful for realizing ultrasensitive probing of chiral information for oriented molecules by plasmon-based nanotechnology.

# I. INTRODUCTION

Chirality, which refers to structures lacking any mirror symmetry planes, is a very intriguing property of molecules. Many biologically active molecules are chiral, which plays a pivotal role in biochemistry and the evolution of life itself.[1-2] Detecting and characterizing chiral enantiomers of these biomolecules are of considerable importance for biomedical diagnostics and pathogen analyses.[3] A common technique for chirality discrimination is CD spectroscopy describing the difference in molecular absorption of left- and right- handed circularly polarized photons.[4-5] In general, the molecular CD signal is typically weak, thus, chiral analyses by such a spectroscopic technique have usually been restricted to analysis at a relatively high concentration.[1-5]

Recent investigations have shown that superchiral fields allow measuring the chiroptical properties of small amounts of molecules with high sensitivity.[6-7] A weak molecular CD signal in the ultraviolet spectral region can be both enhanced and transferred to the visible-near-infrared region when chiral molecules are adsorbed at the surfaces of metallic nanoparticles or in the nanogaps (i.e., hot spots) of particle clusters.[8-29] However, many related discussions assume that the molecules are randomly oriented,[6, 21-24] and the results have been obtained by averaging over the solid angles of the molecular directions. In fact, this rotational degree of freedom for molecules is absent in many cases, i.e. chiral molecules adopt geometries in which they have an axis with a well-defined orientation with respect to the surface of the nanostructures.[9, 26-29] This leads to several orientation-selective signals for appropriately sculpted fields. Although there are also some theoretical researches concentrating on the plasmon-induced CD effect of oriented chiral molecules, they only focus on a fixed number of chiral molecules modeled as point dipoles and the molecular electric quadrupole transitions have also been neglected.[17-20] As for the chiral molecular material with a large collections of molecules, it can be generally regarded as a homogeneous chiral medium being treated with full-field simulations.[24, 30-31] In addition, up to now, all the studies have neglected the molecule electric quadrupolar contribution to the CD.[17, 19, 20-22, 24] When the molecules are randomly oriented, the neglect is reasonable. Nevertheless, if the molecules have defined orientations, the contribution of molecular electric quadrupole transitions to the total CD signal can play an important role.[26, 27, 32-34] For example, the large dissymmetries observed for the experiment proposed by E. Hendry et al cannot be explained by dipolar chiral excitations.[26] The large dissymmetry enhancements may result from the quadrupolar contribution to optical activity. Hence, it is not correct to ignore the molecular electric quadrupolar contribution in the oriented molecular systems.

In this work, we present a rigorous finite element method to study electromagnetic interactions between nanostructures and anisotropic chiral molecular materials with the electric quadrupolar contribution. On the basis of such a method, we calculate the CD in the systems consisting of nanostructures and oriented molecules. We find that the molecular electric dipole-electric quadrupole interaction can play primary roles in the plasmon-induced CD spectroscopy for the system with specific molecule orientations. Furthermore, we have also found that both the quadrupole- and dipole-based CD signals can be amplified tremendously by matching the phases for the electromagnetic fields and their gradients at different regions around the nanostructures, which are occupied by the oriented chiral molecules. Our calculations show that matched phases are as important as the intensities of the electromagnetic field at the molecular positions to enhance the CD signals of oriented molecular medium systems.

This paper is organized as follows: Section 2A presents the effective medium theory of anisotropic chiral molecules based on the semi-classical theory of multipole oscillators induced by electromagnetic fields and the gradients of the electric field. Section 2B shows the method of modeling wave propagation in anisotropic chiral medium. Section 3A is devoted to analyze the CD effects for the symmetric molecule distributions around nanostructures. Section 3B analyzes the CD effects of the asymmetric molecule distributions around nanostructures. Finally, a summary and the main conclusions of this research will be drawn.

**II. Theory and method.**

**A. Effective medium theory for the oriented chiral molecules with electric quadrupole transitions**

Although macroscopic chiral media are composed of microscopic chiral molecules, they are usually regarded as effective media when electromagnetic waves transport in them, which can be described by Maxwell's equations with the following chiral constitutive relations: [35-36]

$$\begin{aligned} \boldsymbol{D} &= \varepsilon_0 \tilde{\varepsilon} \boldsymbol{E} + \tilde{\kappa} \boldsymbol{B} \\ \boldsymbol{B} &= \mu_0 (\boldsymbol{H} - \tilde{\kappa}^T \boldsymbol{E}), \end{aligned} \quad (1)$$

where the superscript '$T$' indicates transposition; $\varepsilon_0$ and $\mu_0$ denote the vacuum permittivity and permeability, respectively. $\tilde{\varepsilon}$ is the relative permittivity dyadic and $\tilde{\kappa}$ is the magnetoelectric dyadic, a key parameter describing the chiral property of the system composed of oriented molecules. In the case of harmonic incident wave, $\tilde{\kappa}$ and $\tilde{\varepsilon}$ are given by the multipole theory purposed by Graham and Raab,[37] and can be expressed as:

$$\begin{aligned} (\tilde{\varepsilon})_{\alpha\beta} &= \varepsilon_b \delta_{\alpha\beta} + (\tilde{\alpha}^{mac})_{\alpha\beta} \\ (\tilde{\kappa})_{\alpha\beta} &= (\tilde{\eta}^{mac})_{\alpha\beta} + i\frac{1}{2}\omega \Lambda_{\beta\gamma\delta} (\tilde{\xi}^{mac})_{\gamma\delta\alpha}. \end{aligned} \quad (2)$$

Here $\Lambda_{\beta\alpha\gamma}$ and $\delta_{\alpha\beta}$ are the Levi-Civita symbol and the Kronecker delta operator; $\varepsilon_b$ is the relative permittivity of the embedded medium, $\omega$ is the angular frequency of the incident wave; $(\tilde{\alpha}^{mac})_{\alpha\beta}$, $(\tilde{\eta}^{mac})_{\alpha\beta}$ and $(\tilde{\xi}^{mac})_{\alpha\beta\gamma}$ denote the corresponding matrix elements of the electric dipole-electric dipole polarizability tensor, electric dipole-magnetic dipole polarizability tensor and electric dipole-electric quadrupole polarizability tensor of macroscopic chiral medium, respectively. In the limit of low density of molecules, the interactions among molecular dipoles can be ignored. Hence, the macroscopic polarizability tensors can be expressed by the polarizability tensors of the single molecule as: [21, 24]

$$(\tilde{\alpha}^{mac})_{\alpha\beta} = n_0 (\tilde{\alpha}^{mic})_{\alpha\beta}$$
$$(\tilde{\eta}^{mac})_{\alpha\beta} = n_0 (\tilde{\eta}^{mic})_{\alpha\beta} \quad (3)$$
$$(\tilde{\xi}^{mac})_{\alpha\beta\gamma} = n_0 (\tilde{\xi}^{mic})_{\alpha\beta\gamma},$$

where $n_0$ is the molecular density; $(\tilde{\alpha}^{mic})_{\alpha\beta}$, $(\tilde{\eta}^{mic})_{\alpha\beta}$ and $(\tilde{\xi}^{mic})_{\alpha\beta\gamma}$ are the corresponding matrix elements of the electric dipole-electric dipole polarizability tensor, electric dipole-magnetic dipole polarizability tensor and electric dipole-electric quadrupole polarizability tensor for a single chiral molecule, respectively. These microscopic polarizability tensors can be derived from the following master equation for quantum states of the single molecule:

$$\hbar \frac{\partial \rho_{nm}}{\partial t} = i \langle n | [\hat{\rho}, \hat{H}_0 + \hat{V}_I] | m \rangle - (\hat{\Gamma} \cdot \hat{\rho})_{nm} \quad (4)$$

where $\hat{\rho}$ is the density matrix, $\rho_{nm}$ ($n$ ($m$)=1, 2) is the corresponding matrix element, $\hat{\Gamma}$ is the operator of the relaxation term which describes the phenomenological damping. The operator $\hat{H}_0$ describes the internal energy of the molecule. $\hat{V}_I$ is the light-molecule interaction operator, which can be written as:

$$\hat{V}_I = -\hat{\mu}\cdot\boldsymbol{E} - \hat{m}\cdot\boldsymbol{B} - \frac{1}{2}\hat{Q}:\nabla\boldsymbol{E} \tag{5}$$

with $\boldsymbol{E} = Re[E_0 e^{-i\omega t}]$ and $\boldsymbol{B} = Re[B_0 e^{-i\omega t}]$ being the incident electric and magnetic fields. $\hat{\mu}$, $\hat{m}$ and $\hat{Q}$ are the molecular electric dipole, magnetic dipole and electric quadrupole operators (the primitive form[38]), respectively. By using the rotating-wave approximation in the linear regime, we can obtain the following solution from Eq. (4) as:

$$\rho_{21} = Re[\frac{\hat{V}_{I_{21}}}{(\omega-\omega_0)\hbar + i\Gamma_{21}}e^{-i\omega t}] \tag{6}$$

With $\hat{V}_{I_{21}} = (-\boldsymbol{\mu}_{21}\cdot\boldsymbol{E} - \boldsymbol{m}_{21}\cdot\boldsymbol{B} - \frac{1}{2}\tilde{Q}_{21}:\nabla\boldsymbol{E})$, where $\mu_{21}$, $m_{21}$ and $Q_{21}$ are the corresponding matrix elements of molecular electric dipole, magnetic dipole and electric quadrupole moments. The parameter $\omega_0$ is the angular frequency of the molecular transition. Then, we can write the electric dipole, magnetic dipole and electric quadrupole moments of the single molecule as:

$$\begin{aligned}\boldsymbol{\mu} &= \boldsymbol{\mu}_{12}(\rho_{21}+\rho_{12})\\ \boldsymbol{m} &= \boldsymbol{m}_{12}\rho_{21}+\boldsymbol{m}_{21}\rho_{12}\\ \tilde{Q} &= \tilde{Q}_{12}(\rho_{21}+\rho_{12}).\end{aligned} \tag{7}$$

Inserting Eqs. (5) and (6) into Eq. (7), we obtain:

$$\begin{aligned}\boldsymbol{\mu} &= \tilde{\alpha}^{mic}\boldsymbol{E} + \tilde{\eta}^{mic}\boldsymbol{B} + \frac{1}{2}\tilde{\xi}^{mic}(\nabla\boldsymbol{E})\\ \boldsymbol{m} &= -\tilde{\eta}^{micT}\boldsymbol{E}\\ \tilde{Q} &= \tilde{\xi}^{micT}\boldsymbol{E},\end{aligned} \tag{8}$$

with

$$(\tilde{\alpha}^{mic})_{\alpha\beta} = -(\frac{1}{\hbar\omega - \hbar\omega_0 + i\Gamma_{12}} - \frac{1}{\hbar\omega + \hbar\omega_0 + i\Gamma_{21}})(\boldsymbol{\mu}_{12} \otimes \boldsymbol{\mu}_{21})_{\alpha\beta}$$

$$(\tilde{\eta}^{mic})_{\alpha\beta} = -(\frac{1}{\hbar\omega - \hbar\omega_0 + i\Gamma_{12}} + \frac{1}{\hbar\omega + \hbar\omega_0 + i\Gamma_{21}})(\boldsymbol{\mu}_{12} \otimes \boldsymbol{m}_{21})_{\alpha\beta} \qquad (9)$$

$$(\tilde{\xi}^{mic})_{\alpha\beta\gamma} = -(\frac{1}{\hbar\omega - \hbar\omega_0 + i\Gamma_{12}} - \frac{1}{\hbar\omega + \hbar\omega_0 + i\Gamma_{21}})(\boldsymbol{\mu}_{21})_{\alpha}(\tilde{Q}_{21})_{\beta\gamma}$$

From Eqs. (2), (3) and (9), the general forms for the response tensors of anisotropic chiral medium with the electric quadrupole transitions can be obtained.

**B. The method for modeling wave propagation in the anisotropic chiral medium**

A number of algorithms have been developed to calculate transport properties of the wave in the media.[39-40] One of them is the finite element method (FEM) to solve the full Maxwell's equations. Over the last decades, numerical modeling algorithms are now being built into commercial or open-source software. COMSOL Multiphysics software,[41] which is based on solution of second order differential equations for the electric field, is a numerical simulator based on the FEM. Recently, a rigorous finite element method that takes both the complex plasmonic structure and isotropic chiral medium into account has just been presented by using a modified COMSOL RF simulation environment.[24] But an effective numerical approach to study the oriented molecular systems has not been elaborated. Here, we consider the molecular orientation-effects and the non-ignorable quadrupolar term in light-molecule interactions. The oriented molecule materials are regarded as anisotropic chiral media. The Maxwell's equations for wave propagations in these chiral media can be expressed as:

$$\nabla \times \nabla \times \boldsymbol{E} = \omega^2 \mu_0 \varepsilon_0 \tilde{\varepsilon} \boldsymbol{E} + i\omega\mu_0 \tilde{\kappa}(\nabla \times \boldsymbol{E}) + i\omega\mu_0 (\nabla \times \tilde{\kappa}^T) \cdot \boldsymbol{E} - i\omega\mu_0 \tilde{\kappa}^T \stackrel{\cdot}{\times} (\boldsymbol{E}\nabla). \qquad (10)$$

Here $\tilde{A} \overset{\cdot}{\underset{\times}{}} \tilde{B} = A_{ij} e_i e_j \overset{\cdot}{\underset{\times}{}} B_{ks} e_k e_s = A_{ij} B_{ks} (e_i \times e_s)(e_j \cdot e_k)$. If we assume that the magnetoelectric dyadic is space-independent, the above equation can be reduced to:

$$\nabla \times \nabla \times \boldsymbol{E} = \omega^2 \mu_0 \varepsilon_0 \tilde{\varepsilon} \boldsymbol{E} + i\omega \tilde{\kappa} \mu_0 (\nabla \times \boldsymbol{E}) \\ -i\omega \mu_0 [(\tilde{\kappa}_{jy} \nabla_z E_j - \tilde{\kappa}_{jz} \nabla_y E_j) \boldsymbol{e}_x + (\tilde{\kappa}_{jz} \nabla_x E_j - \tilde{\kappa}_{jx} \nabla_z E_j) \boldsymbol{e}_y + (\tilde{\kappa}_{jx} \nabla_y E_j - \tilde{\kappa}_{jy} \nabla_x E_j) \boldsymbol{e}_z]. \tag{11}$$

In order to model wave propagations in the anisotropic chiral medium, we need to replace the wave equation for usual media by that of chiral media in Eq. 11.[24, 30] Meanwhile, all parameters should be redefined through $\boldsymbol{E}$ according to the chiral constitutive relations described by Eq. 1. For example, the magnetic field vector should be expressed as:

$$\boldsymbol{H} = \frac{(\nabla \times \boldsymbol{E} - i\mu_0 \omega \tilde{\kappa}^T \boldsymbol{E})}{-i\omega \mu_0}. \tag{12}$$

In addition, boundary conditions for Eq. 11 are not trivial, although they follow from the usual boundary conditions with:

$$\boldsymbol{n} \times (\boldsymbol{H}_2 - \boldsymbol{H}_1) = 0 \\ \boldsymbol{n} \times (\boldsymbol{E}_2 - \boldsymbol{E}_1) = 0. \tag{13}$$

Combing Eq. 13 with the redefined magnetic field vector in Eq. 12, the surface current on the interface can be derived as:

$$\boldsymbol{n} \times (\frac{\boldsymbol{B}_1}{\mu_0} - \frac{\boldsymbol{B}_2}{\mu_0}) = \boldsymbol{n} \times (\tilde{\kappa}^T \boldsymbol{E}) = \boldsymbol{J}_b, \tag{14}$$

where $\boldsymbol{n}$ is a vector normal to the boundary of the chiral medium (we only consider one kind of chiral medium in our system). What we need to emphasize is that the boundary conditions are indeed trivial (see in Eq. 13) for the bianisotropic media.[42] There are other methods to simulate the chiral medium in Comsol. In fact, it is possible to leave the weak form of the wave equation in Comsol unchanged, and only to change the constitutive equations in "Equation View" of the "Wave Equation". Comparing the present case with that of the isotropic chiral medium,[24, 30] the response scalars have been replaced by tensors with the non-ignorable quadrupolar term in light-molecule interactions. In addition, the chirality parameter $\tilde{\kappa}$ and dielectric tensors $\tilde{\varepsilon}$ for the oriented chiral molecular medium are derived from the first-principles calculations, shown in Section 2A. Both the ED-MD and ED-EQ interactions are considered during the process of solving the master equation for quantum states of the single oriented molecule. Consequently, the molecular electric quadrupolar contribution to the CD signal, which was always neglected in the past studies[17, 19, 20-22, 24], can be calculated based on our method. The proposed method can be used to investigate the CD effect of the systems composed of more complex nanostructures and oriented chiral molecular medium of arbitrary shape with both ED-MD and ED-EQ interactions.

### III. NUMERICAL RESULTS AND DISCUSSION.

**A. CD effect of symmetric molecule distributions around nanostructures.**

We first consider the oriented molecular systems without nanostructures. The schematics of the systems with different orientations of the molecular electric dipole moments are shown in the insets of Figure. 1. The chiral medium strips, which are arranged periodically with period $p=320$ nm, are infinite in the z direction, 0.75 nm and 10 nm along x- and y-axis (the size of the chiral medium is suited for the nanostructures discussed below). Two strips in each period are 80 nm apart. Throughout this work, the convenient parameters for the molecular electric dipole (ED) and magnetic dipole (MD) moments are $\mu_{12} = |e| r_{12}$ and $m_{12} = 0.5 i |e| r_0 \omega_0 r_{12}$. The form of the molecular electric quadrupole (EQ) moments can be written as: $Q_{21,ij} = 0.5 |e| r_q^2$ ($i \neq j$) and

$Q_{21,ii} = \pm |e| r_q^2$, where the subscripts $i$ and $j$ represent the coordinate components x, y and z, respectively, and $|e|$ is the absolute value of the electric charge. The form of the molecular EQ moments will be changed with the rotation of the molecular ED moments. In this condition, minus sign is only used for the diagonal term of the EQ operators ($Q_{21,ii}$) when the orientations of molecular ED moments ($\mu_{12}$) are along the $i$-axis. In the following calculations, we choose parameters:[19] $r_{12} = 0.2$ nm, $r_0 = 0.005$ nm, $r_q = 0.025$ nm and $\Gamma_{21} = 0.3$ eV; The wavelength of the molecular transition is taken as $\lambda_0 = 2\pi c/\omega_0 = 300$ nm; The molecule density is $n_0 = (2 \text{ nm})^{-3}$ and the relative permittivity of the embedded medium is $\varepsilon_b = 1$. Figure 1(a) and (b) show the numerical results for the difference in molecular absorptivity (1-R-T, R and T are the reflectance and transmittance of the system) for left- and right- handed circularly polarized photons when the orientations of molecular ED moments are along x- and z-axis, respectively. Based on the numerical validation, we find that the influences of higher diffraction orders on the CD signal are negligible with the wavelength of the incident light larger than 200 nm in all of the following systems. We can see that if we neglect the molecular EQ transitions, the CD signal will decrease by increasing the angle ($\theta$, shown in Figure 1) between the molecular electric and magnetic dipole moments (black and green lines correspond to $\theta = 0$ deg and $\theta = 45$ deg) and equals to zero with $\theta = 90$ deg (blue line). Here, the molecular MD moments are rotated in x-y and x-z planes, respectively. As for the CD signals resulting from the molecular ED-EQ interaction (red lines), obviously, the molecular EQ transitions contribute in the same order as the MD on the whole CD for the oriented molecular systems.[43] However, when the orientations of the molecular ED moments are parallel to the wave vector for the incident light, the CD signals nearly vanish as shown in Figure 1(c). On this occasion, the absorption of the molecular system becomes rather small, which results the corresponding CD signals can also be neglected. For comparison, we also plot the CD signal for the isotropic chiral medium with orange lines in Figure 1. It can be seen that the oriented molecular chiral medium may exhibit larger optical activity than the isotropic molecular system.

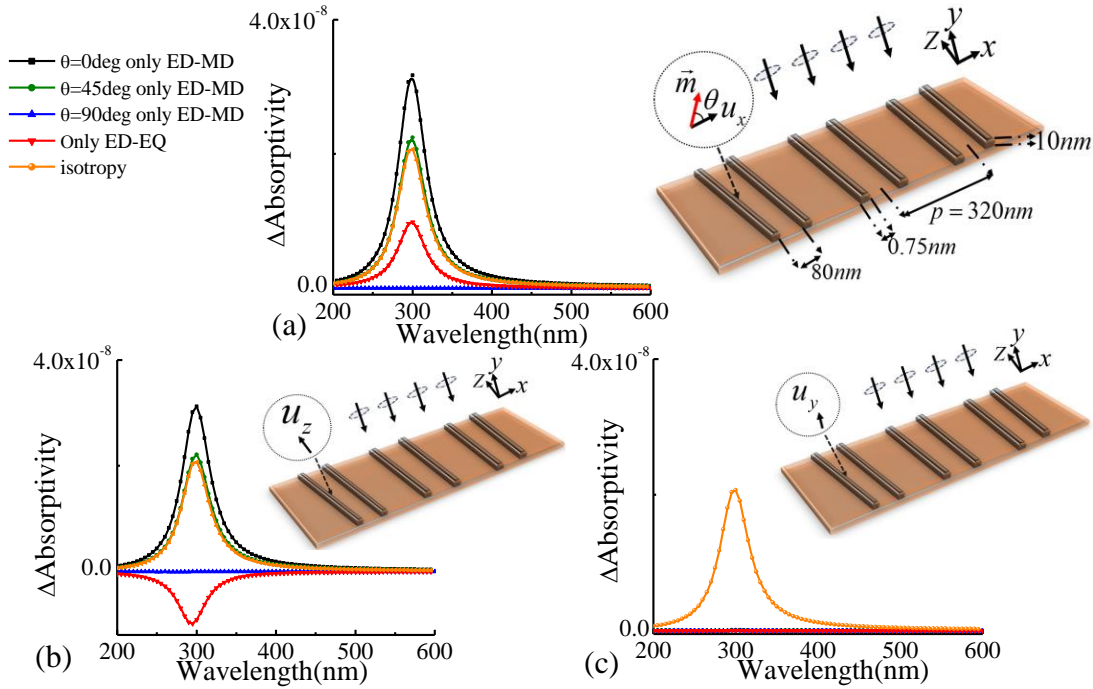

**Figure 1.** (Color online) The difference in absorptivity of left- and right- handed circularly polarized photons excitation for the molecular systems that the chiral medium strips, arranged periodically with period $p = 320$ nm, are infinite in the z direction and 0.75 nm (10 nm) along x(y)- axis. Two strips in each period are 80 nm apart. The wave vector of the incident light is along the y-axis (identical with following systems). The orientations of ED moments are along (a) x-axis, (b) z-axis and (c) y-axis. The black, green and blue lines correspond to the CD signals, without EQ term, with the angles between the molecular ED and MD moments equaling to 0deg, 45deg and 90deg, respectively. The red lines represent the result of the EQ contribution. The orange lines show the solution of the isotropic chiral medium.

Some metallic nanostructures present a unique property, known as localized surface plasmon resonances (SPRs), under resonance excitations by external fields.[44-45] The excitation of SPRs produces intense electric field enhancement in close proximity to the surfaces of metallic nanostructures, which promotes the light-matter interactions for the molecules located in the near field of nanostructures, leading to various plasmon-enhanced spectroscopies. In the following, we will investigate CD spectra of a few complex systems formed by the oriented chiral molecules and metallic nanostructures. As shown in the insets of Figure 2, the molecule-metal hybrids are arranged in an array of period $p$=320 nm. The chiral medium strips are infinite in the z direction, 0.75 nm and 10 nm along x- and y-axis, respectively. As for the gold nanoribbon (infinite along z-axis), the sizes in x and y directions are 80 nm and 10 nm. For the dielectric functions of gold, Johnson's data were adopted.[46] The orientations of the molecular ED moments are presented in the inset of Figure 2. Comparing with the case of bare chiral molecules (Figure 1), we found that the plasmon-induced CD signal appears in these complex systems when the local plasmonic resonance is excited (610 nm). As shown in Figure 2(a)-(c), we found that both the molecular ED-MD (black line) and ED-EQ (red line) interactions have a significant role on the plasmon-induced CD signal with different orientations of molecular ED moments. Meanwhile, when the molecules are achiral, the corresponding CD signals become 0 (blue line). Moreover, we need to emphasize that the above results do not depend on the orientations of the molecular MD moments. The averaged CD spectrum induced by the MD transitions (black line in Figure. 2(d)) is equal to the one when chiral media are considered to be isotropic (orange line in Figure. 2(d)). And the averaged CD signal resulting from the molecular EQ transitions is disappeared (red line in Figure. 2(d)), which supports the conclusion that EQ contribution to the differential absorption averages to zero in isotropic samples.[6, 24]

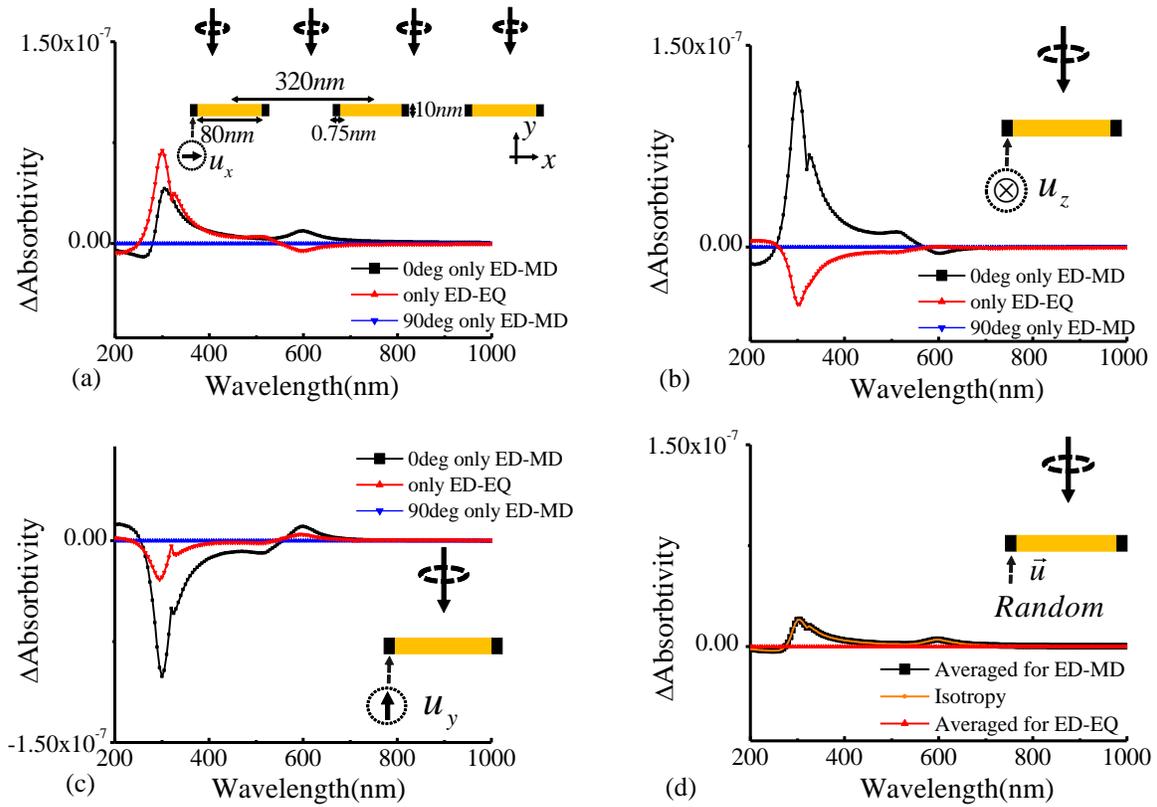

**Figure 2.** (Color online) The difference in the absorptivity of left- and right- handed circularly polarized photons excitation for the molecular systems formed by the oriented molecules and gold nanoribbons. The orientations of the molecular ED moments are along (a) x-axis (b) z-axis and (c) y-axis. The black and blue lines correspond to the CD signal, without EQ term, with the angle between the molecular ED and MD moments equaling to 0deg and 90deg, respectively. The red lines represent the result for the EQ contribution. (d) The comparison between the averaged CD signals, the ED-MD interaction (black) and ED-EQ interaction (red), and the CD effect of isotropic chiral medium system (orange).

As shown in the above results, the molecular ED-EQ interaction can play an equally important role as EQ-MD interaction on the plasmon-induced CD effects of oriented molecular systems. Next, we will demonstrate that the molecular EQ contribution can play a crucial role on the plasmon-induced CD signal when chiral medium strips are put in the nanogaps. The corresponding schematics are presented in the insets of Figure 3. The dimensions of the gold nanoribbons are the same to the models as described in Figure 2. And the lengths along x- and y-axis of the chiral medium strips are 1.5 nm and 10 nm, respectively. The gap size (x direction) is 3 nm, and the period is also chosen to be 320 nm. It is shown clearly that the contributions of the molecular ED-MD interaction play a key role in the plasmon-induced circular dichroism spectra (the resonance is excited at 700 nm), and the contribution of the quadrupolar term almost disappears when the orientations of molecular ED moments are along x-axis, shown in Figure 3. However, if we alter the orientations of the molecular ED moments along z or y-axis, the situation is completely opposite. In this case, the molecular EQ transition plays a crucial role in the plasmon-induced CD spectra, this can be seen clearly in Figure 3(b) and (c). On the other hand, the averaged CD signals display the same behaviors as the above single nanoribbon system, shown in Figure 3(d), although the EQ transitions may play a more important role than the MD term on the CD spectra for the oriented molecules.

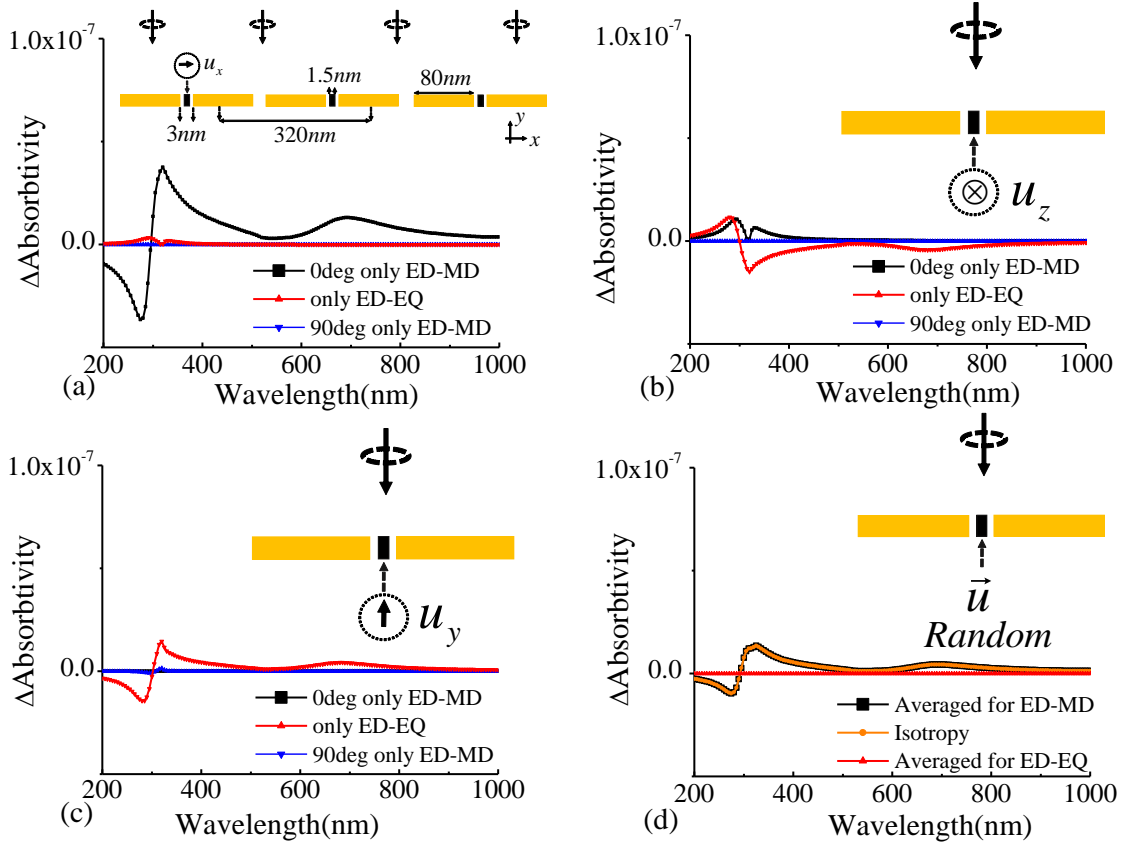

**Figure 3.** (Color online) The difference in the absorptivity of left- and right- handed circularly polarized photons excitation for the molecular systems formed by the oriented molecules and gap nanoribbons.

From the above results, we can see that the plasmon-induced CD is sensitive to the orientations of the molecules. These obvious oriented phenomena are mainly due to the remarkable anisotropy for the electromagnetic environment at the molecular positions produced by the metallic nanostructures, which has been shown in Figure 4. Figure 4(a)-(c) show the averaged intensities of the electric fields, magnetic fields and gradients of electric field within the volume $V_{blue}$ (blue areas in the inset of Figure 4(a)), occupied by the chiral medium strips (show in Figure 2), as functions of the wavelength of the incident left-hand circular polarized light, respectively. The power of the incident light is taken as 1W/m in each period. For example, the averaged value for the x-polarized electric field can be expressed as: $<|E_x|> = \int_{V_{blue}} |E_x| dV / V_{blue}$. In Figure 4(a) and (b), the black, red and blue lines represent the polarizations of the electromagnetic field along x, y and z directions, respectively. And for Figure 4(c), these lines illustrate the gradients of the electric field along x, y and z-axis. It is seen clearly that both the fields and their gradients exhibit the characteristics of anisotropy. As shown in Figure 4(a), we find that the averaged intensity of the electric field with the polarization direction along the x-axis is 10.8 and 9.8 times larger than that along y- and z-axis, respectively, when the local plasmonic resonance was excited (610 nm). The anisotropy of the electric fields at the positions of the chiral medium causes directional-selectivity for the ED moments during the process of light-molecules interaction. Similarly, the anisotropies for the magnetic fields (shown in Figure 4(b)) and gradients of electric field (shown in Figure 4(c)) cause directional-selectivity for the MD and EQ moments, respectively. Such a phenomenon related to the molecular orientations can be seen more clearly in the nanogap system. Figure 4(d)-(f) show the averaged intensities of the electric fields, magnetic fields and gradients of the electric field within the volume $V_{blue}$ (blue areas in the inset of Figure 4(d)), occupied by the chiral medium strips (show in Figure 3), as functions of the wavelength of the incident left-hand circular polarized light, respectively. For comparative purposes, the power of the incident light is also chosen to be 1 W/m in each period. In such a case, the averaged intensity of the x-polarized electric field is 84.8 and 32.0 times larger than that polarized along y- and z-axis, respectively, when the local plasmonic resonance happens (700 nm). And comparing with the case of the single nanoribbon system, the averaged intensity of the x-polarized electric field is also amplified nearly 3-fold. From the above results, we can see that the value of x-polarized field is much larger than y- and z-polarized fields at molecular positions in the gap nanoribbon system. However, the magnetic fields and gradients of electric field have not been enhanced simultaneously compared to the single nanoribbon system, shown in Figure 4(d) and (e). The gradients of the electric field along x- and y-axis are only one third of the corresponding values of the single nanoribbon system. Consequently, we can speculate that the plasmonic CD signals induced by ED-MD and ED-EQ interactions are both dominated by the x-polarized electric fields in the nanogap system. As illustrated in Figure 3, the ED-MD interaction can induced large CD signals only when the orientations of the ED moments are along x-axis. Because only in this case, the coupling of the x-polarized electric field to ED-MD transitions can occur. As for the case of ED-EQ interaction, the large CD signals can be induced when the orientations of the ED moments are along y- or z-axis with the coupling of the x-polarized electric field to ED-EQ transitions existing. These phenomena are obviously different from the cases of the single nanoribbon system that the CD effect produced by different orientations of molecules are all significant, shown in Figure 2. It can be explained by the fact that both the electric fields and gradients of electric field all have non-ignorable values in the single nanoribbon system, shown in Figure 4(a)-(c). Finally, we need to emphasize that the comparison between the results derived from the two nanostructures, shown in Figures 2 and 3, is not to distinguish the values of the CD signals. Our purpose is to disclose the physical origin of the phenomena that the dipole- and quadrupole-based CD signals have an important relationship with the molecular orientations.

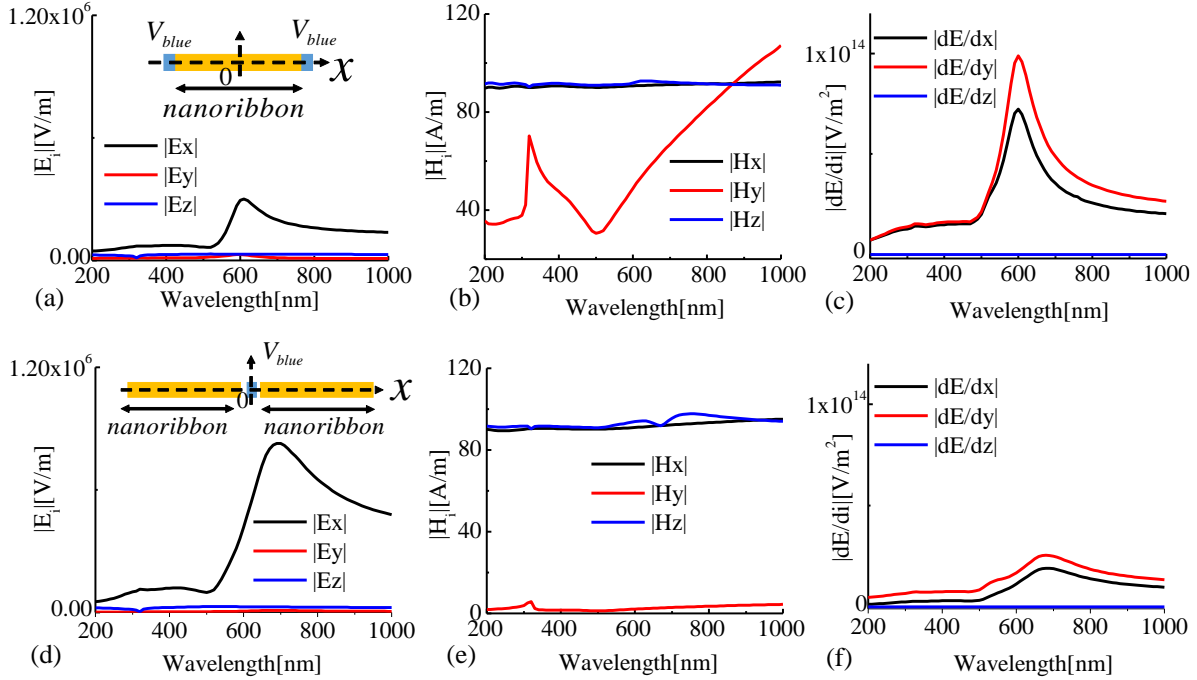

**Figure 4.** (Color online) The averaged intensity for the (a) electric fields, (b) magnetic fields and (c) gradients of the electric field within the volume $V_{blue}$ (blue areas in the inset of Figure 4(a)), occupied by the chiral medium strips, as a function of the wavelength of the periodic single-nanoribbons system. (d)-(f) plot the corresponding components for the periodic gap nanoribbons system.

## B. CD effect of asymmetric molecule distributions around nanostructures.

The above investigations only focus on the symmetric distributions of molecules around nanostructures. In fact, when the symmetry of the molecule distributions is broken, more prominent phenomenon can appear. In the following, we will investigate the CD of asymmetric distributions of the oriented molecular systems, which are presented in the insets of Figure 5. The chiral molecular media are only put at one side of the gold nanoribbons that the dimensions along x- and y-axis are 1.5 nm and 10 nm, respectively. As shown in Figure 5(a)-(f), the dipole-based CD spectra (black lines) are nearly identical with the corresponding symmetric systems (black lines in Figure 2(a)-(c)) when the orientations of the molecular ED and MD moments are paralleled. What may come as a surprise is that there are still non-ignorable CD signals (green lines) even the molecules are achiral, when the orientations of the molecular ED moments are along x- or y-axis, presented in Figure 5(a)-(d). Especially, when the molecular ED and MD moments are parallel to the y- and x-axis, the corresponding dipole-based CD peak (green lines in Figure 5(c)-(d)) is amplified tremendously comparing with the case that molecules were symmetrically distributed (black line in Figure 2(c)). However, when we change the directions of the molecular ED moments along the z-axis, shown in Figure 5(e) and (f), the ED-MD induced CD effects disappear if the molecules are achiral. In addition, the ED-EQ induced CD peaks (red lines) is nearly hundreds-fold over the corresponding symmetric system (red lines in Figure 2(b) and (c)), when the orientations of the molecular ED moments are along y- or z-axis, presented in Figure 5(c)-(f). On the other hand, in many case, the plasmon-induced CD signals are entirely different, when the molecules are positioned at different sides of the gold nanoribbon. Moreover, CD signals with asymmetric molecule distributions also depend on the orientations of the molecular MD moments, show in Figure 5. The above phenomena resulting from molecular asymmetric distributions can also exist in the nanogap system with the molecules clinging to the left or right side in the gap region, shown in the insets of Figure 6. From Figure 6(a)-(d), we can see that there are non-ignorable CD signals (green lines), even the molecules are achiral. However, when the directions of the molecular ED moments are along the z-axis, the corresponding CD signal will disappear with the molecules being achiral, shown in Figure 6(e) and (f). In this condition, the ED-EQ transition dominates the whole plasmon-induced CD signal, and the CD-peak is nearly hundreds-fold over the corresponding symmetric system (red line in Figure 3(b)).

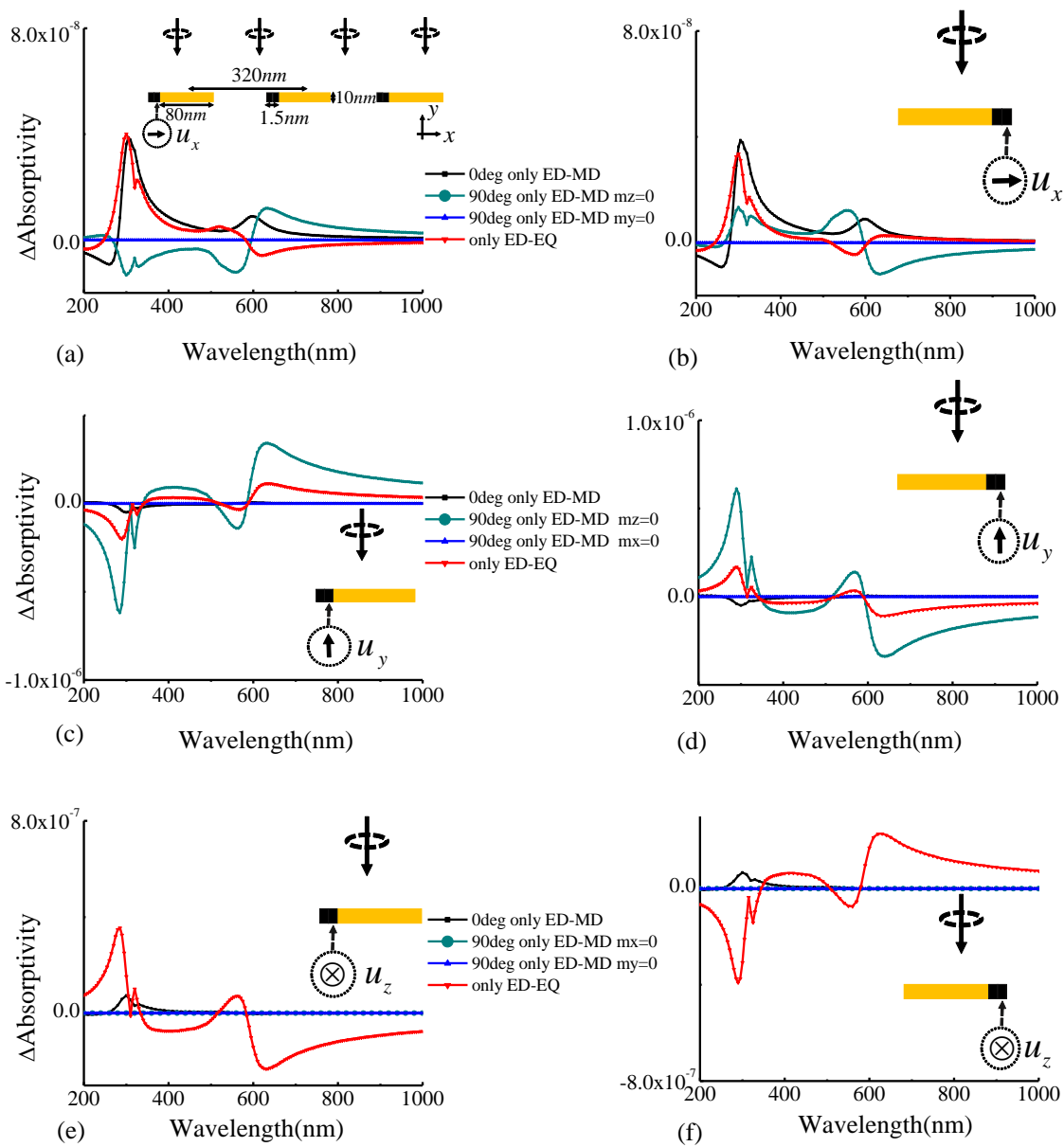

**Figure 5.** (Color online) The difference of the absorptivity of left- and right- handed circularly polarized photons excitation with the asymmetric molecule distributions formed by the oriented molecules and gold nanoribbons. (a), (c) and (e) left side, (b), (d) and (f) right side exist molecules with the orientations of the molecular ED moments along x-axis, y-axis and z-axis, respectively. The black lines correspond to the CD signals, without EQ term, with the angle between the molecular ED and MD moments equaling to 0deg. The green and blue lines correspond to the CD signals of achiral molecules, which are not take the ED-EQ

interaction into account, with different components of MD moments equaling to 0. The red lines represent the results for the EQ contributions.

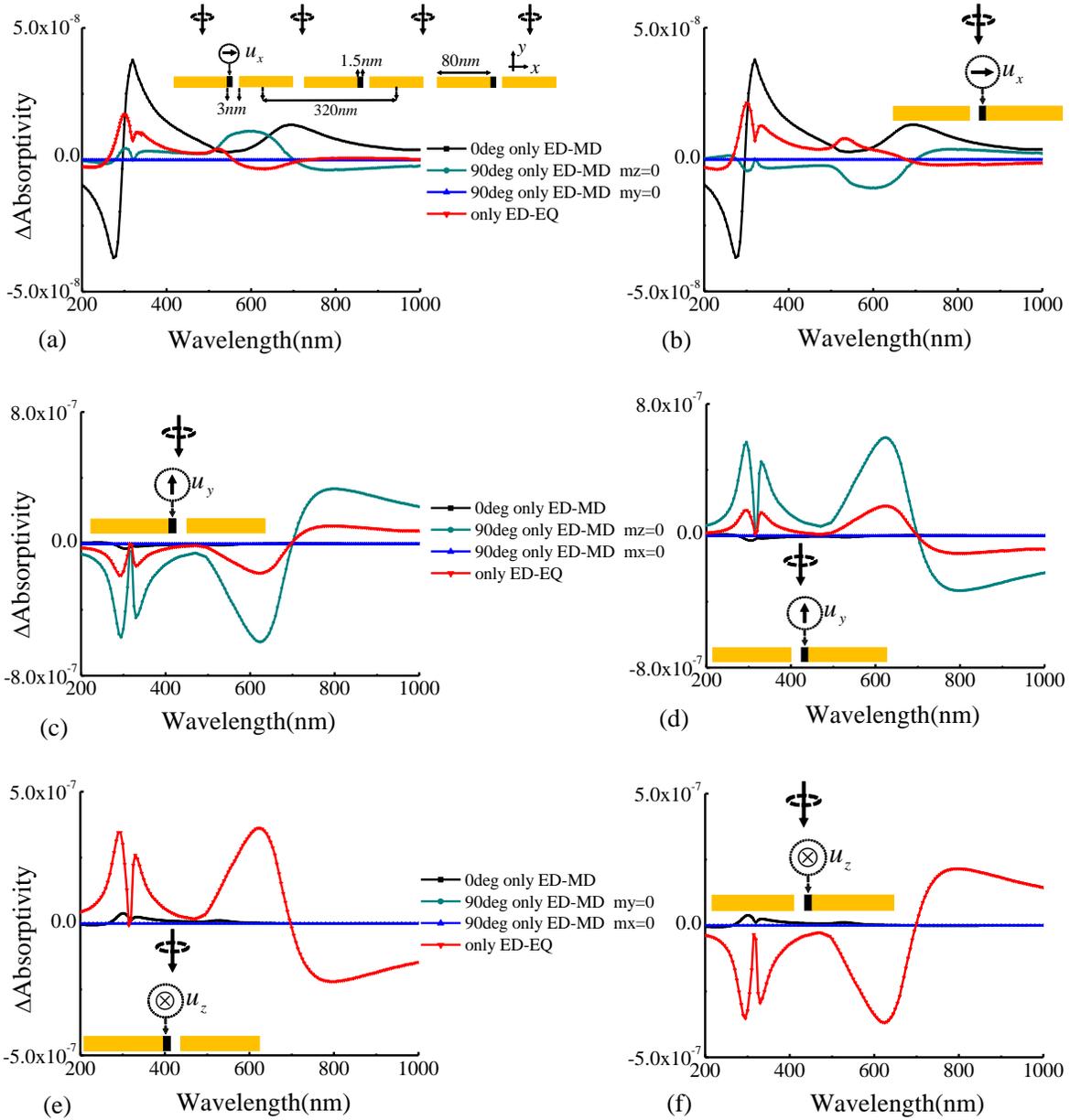

**Figure 6.** (Color online) The difference of the absorptivity of left- and right- handed circularly polarized photons excitation with the asymmetric molecule distributions formed by the oriented molecules and gap nanoribbons.

To disclose the physical origin of the phenomena happened with asymmetric molecular distributions, we plot the differences of averaged phases for near-fields between different sides of the nanostructures as functions of the wavelength for the incident left-hand circular polarized light, shown in Figure 7. Figure 7(a)-(c) show differences of averaged phases of electric fields, magnetic fields and gradients of electric field between $V_{red}$ (red areas in the inset of Figure 7(a)) and $V_{blue}$ (blue areas in the inset of Figure 7(a)) areas around the single nanoribbon system as functions of the wavelength for the incident left-hand circular polarized light, respectively (the difference of averaged phase of x-polarized electric fields between $V_{red}$ and $V_{blue}$ areas can be expressed as: $<\Delta Arg(E_x)> = \int_{V_{red}} Abs[Arg(E_x)]dV/V_{red} - \int_{V_{blue}} Abs[Arg(E_x)]dV/V_{blue}$, with the variation range of the corresponding phases being $-\pi \to \pi$). In Figure 7(a) and (b), the black, red and blue lines represent the polarization of the electromagnetic fields along x, y and z directions, respectively. In Figure 7(c) these lines illustrate the electric field gradients along x, y and z-axis. We find that the difference of averaged phases between both sides of the single nanoribbon are nonzero for the y-polarized electric and magnetic fields and gradients of electric field along x-axis. Due to these non-uniform phase distributions for the electromagnetic components, which are coupled to the molecular ED-MD or ED-EQ transition, the oriented chiral molecular media located at different sides of the nanostructures may induced CD signals with the opposite sign, even mutual offset (green lines in Figure 5(c) and (d)). For example, comparing with the case that molecules are symmetrically distributed (Figure 2(b)), asymmetric molecular distributions (Figure 5(e) and (f)) may enlarge the CD-peak (red lines) induced by ED-EQ interaction over hundreds-fold. This can be explained by the condition that the gradients of the electric field along x-axis are coupled with molecular ED-EQ transitions when the orientation of the molecular electric dipole moment is along z-axis. This will result in the CD signals produced by the chiral molecular media, which lie in different sides of the single nanoribbon, possessing opposite sign (due to non-uniform sign of the averaged phase of the gradients of the electric field along x-axis in different sides of the single nanoribbon). Similar results also exist in the gap nanoribbon systems. Figure 7(d)-(f) show differences of averaged phases of electric fields, magnetic fields and gradients of electric field between $V_{red}$ (red areas in the inset of Figure 7(d)) and $V_{blue}$ (blue areas in the inset of Figure 7(d)) areas around the gap nanoribbon system as functions of the wavelength of the incident left-hand circular polarized light, respectively. We find that the nonzero components of differences of averaged phases are identical with the single nanoribbon system. Hence, similar phenomena happened with asymmetric molecular distributions for single nanoribbon and gap nanoribbon systems. Our calculations show that matching the phase for the electromagnetic components, which are coupling with molecular ED-MD or ED-EQ interaction, is significant for the CD enhancement of oriented chiral molecules system. Finally, we need to clarify that the phases discussed here are not the local phase difference between the electric and magnetic fields. What we concerned is the spatial distributions of the phases for the specific electromagnetic fields, which are coupled with ED-MD or ED-EQ interaction of the oriented chiral molecules to exhibit chiroptical effects.

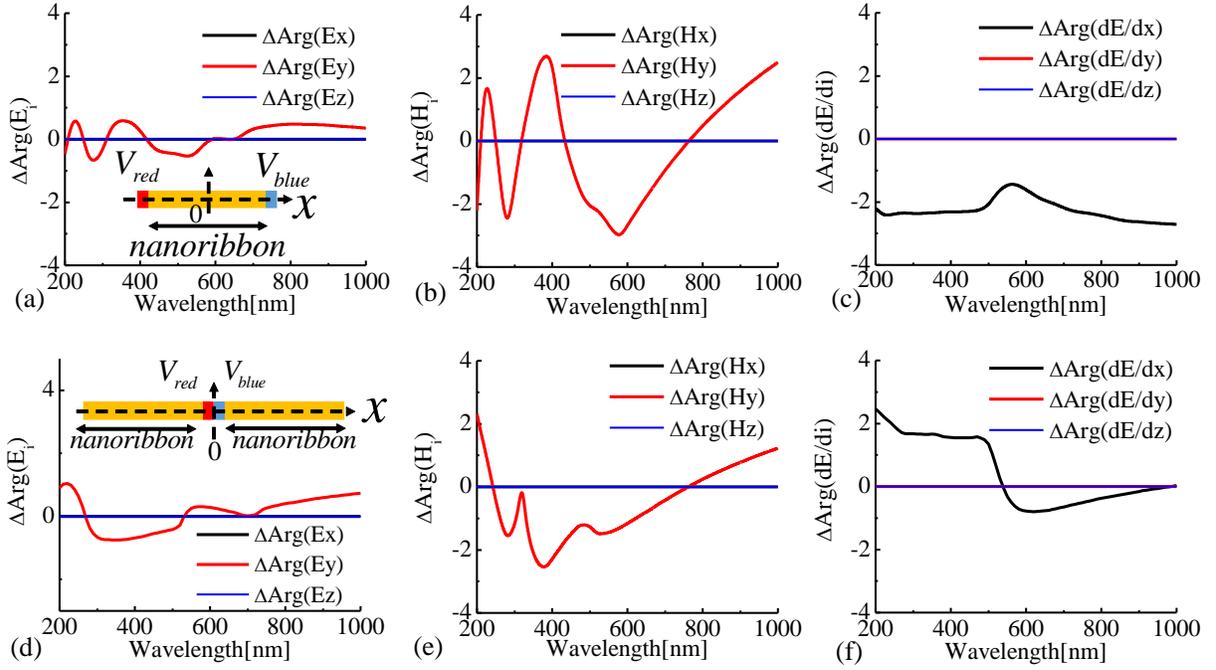

**Figure 7.** (Color online) The difference of the averaged phases for the (a) electric fields, (b) magnetic fields and (c) gradients of the electric fields between $V_{red}$ (red areas in the inset of Figure 7(a)) and $V_{blue}$ (blue areas in the inset of Figure 7(a)) areas around the single nanoribbon system as a function of wavelength of the incident left-hand circular polarized light. (d)-(f) plot the corresponding components for the periodic gap nanoribbons system.

**Conclusion.**

We have used a rigorous finite element method to study the CD effect of various systems with spatial symmetric and asymmetric molecular distributions around nanostructures. The orientation-sensitive CD effects are investigated with non-ignorable EQ term in light-molecule interaction, which was always omitted under the assumption of a large collections of molecules random oriented. Although the CD signal resulting from the molecular quadrupole moments in our system can play a key role in many cases, the relatively low enhancements of near field gradients proposed by our one-dimensional wires systems still cannot amplify the CD above the experimental detection limit (the same as the case of the isotropic chiral medium).[24] By designing nanostructures with steeper field gradients, the much larger CD signal resulting from molecular quadrupole transition may be achieved. Furthermore, we have demonstrated that both the quadrupole- and dipole-based CD signals can be improved greatly by laying the chiral molecules asymmetrically around the nanostructures to match the phases for the electromagnetic components, which are coupled with molecular ED-MD or ED-EQ interaction, at the molecule positions. This method can lead to the uniform sign of the CD produced by the oriented chiral molecules, which are located at different spatial positions near the nanostructures. We believe that these findings would be helpful for realizing ultrasensitive probing of chiral information for oriented molecules by plasmon-based nanotechnology.

**AUTHOR INFORMATION**

**Corresponding Author**

* E-mail: zhangxd@bit.edu.cn

**Notes**

The authors declare no competing financial interest.

**Acknowledgment**

This work was supported by the National Natural Science Foundation of China (Grant No. 11274042) and the National Key Basic Research Special Foundation of China under Grant 2013CB632704.

**Supporting Information**

Details of analytical derivations for the anisotropic wave equation and the check of our numerical method are given. This material is available free of charge via the Internet at http://pubs.acs.org.

**TOC Graphic:**

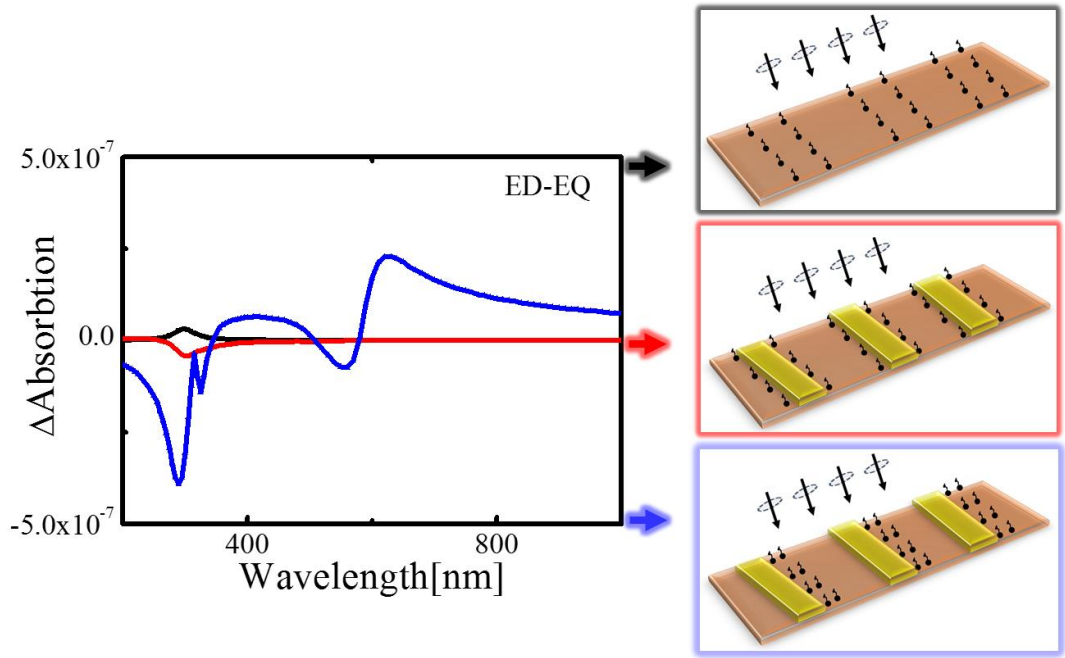